# Enhanced Sensitivity of THz NbN Hot Electron Bolometer Mixers

B. Mirzaei, J. R. G. Silva, W. J. Vreeling, W. Laauwen, D. Ren, and J. R. Gao

*Abstract*— We studied the effect of the NbN/Au contact on the sensitivities of a NbN hot electron bolometer (HEB) mixer by measuring the double sideband (DSB) receiver noise temperature ($T_{rec}^{DSB}$) at three local oscillator frequencies of 1.6, 2.5 and 5.3 THz. The HEB has cleaned contact structures with a thick Au layer. We demonstrated low mixer noise temperatures ($T_{mixer}^{DSB}$) of 240 K and 290 K at 1.6 and 2.5 THz, respectively. The latter reach roughly 3 times the quantum noise at their frequencies. The mixer is developed for the proposed OASIS and SALTUS (concept) missions. The enhanced $T_{mixer}^{DSB}$ are more than 30 % better in comparison with published NbN HEB mixers. The improvement can reduce the integration time of a heterodyne instrument roughly by a factor of 2. The $T_{mixer}^{DSB}$ of the same HEB has shown limited improvement at 5.3 THz, which is partly due to non-optimized antenna geometry. Besides, the results also help to understand device physics of a wide HEB (4 μm) at high frequencies.

## INTRODUCTION

HOT electron bolometer (HEB) mixers based on a thin NbN superconducting bridge are so far the most sensitive heterodyne detectors for high spectral resolution ($\geq 10^6$) spectroscopy in astronomic observations in the frequency range between 1 and 6 THz. Due to water vapor absorption in the atmosphere, such observations can only be realized through air-, balloon-, and space-borne observatories. NbN HEBs have been flown, for example, on HIFI-Herschel [1], SOFIA [2], and STO2 [3], and will be flown on GUSTO [4] and ASTHROS [5]. They are also the choice for FIRSS (or LETO) [6], OASIS [7], and SALTUS [8] space missions, proposed (or to be proposed) to ESA or NASA. In particular, due to limited lifetime of a space mission, a low mixer noise temperature ($T_{mixer}^{DSB}$) is highly demanded in order to make optimal use of the observation time since the integration time of the heterodyne instruments is proportional to the square of the noise temperature [9].

Many years of research and development at different research groups in the world have been devoted to realizing, understanding, and improving low noise NbN HEBs [10-12]. Here we only quote the sensitivities for one particular frequency, namely receiver noise temperatures ($T_{rec}^{DSB}$, the sensitivity of a mixer including the contribution from all the optical components) measured at 1.6 THz or extrapolated to this frequency, which are, for example, 750 K from HIFI [1], 690 K from STO2 [13], and 760 K from GUSTO [14]. The $T_{rec}^{DSB}$ have not been reduced in the past decade. The HEBs for STO2 [13] and GUSTO [14] were fabricated at TU Delft/SRON using contacts that have a superconducting interlayer of either NbTiN [15] or Nb [16] between thin Au and ultra-thin NbN layers. This interlayer helps to overcome the proximity effect to keep the NbN underneath the Au being superconductive. Thus, it also prevents the reduction of the critical temperature ($T_c$) of the NbN bridge. However, the interlayer may introduce side effects, e.g., on RF loss, which can affect $T_{mixer}^{DSB}$. Here we present the results of a new NbN HEB mixer using contacts that were cleaned and have a direct interface between thick Au and NbN. We show that the DSB mixer noise temperature improves by roughly 30 % at 1.6 and 2.5 THz compared to the results that we have previously reported or that are in the literature.

The aim of this study is also to demonstrate sensitive HEB mixers for the SALTUS mission [8], which requires three arrays of HEBs, operating respectively in the 1-2 THz frequency band for numerous molecules, such as $H_2O$ lines, at a single frequency of 2.67 THz for HD (1-0) line, and at a single frequency of 5.33 THz for HD (2-1) line.

## II. HEB GEOMETRY AND FABRICATION

The HEB under study (labeled as "OASIS BM2 7B") has a 400 nm long and 4 μm wide NbN bridge, with a normal state resistance ($R_N$) of 79 Ω and a $T_c$ of 8.8 K, embedded in the center of a spiral antenna. We choose the large size instead of a typical size of 200 nm × 2 μm NbN bridge as used in [17, 18] because of, as to be discussed, the strong lateral proximity effect due to the contacts. Our fabrication method could bring the $T_c$ to a lower value that can affect the LO power and the operating temperature. We aim for an LO power of around 200 nW, which is desirable in view of the receiver stability and availability of the power from existing LO sources. In addition, we expect wider HEBs to show a lower contact resistance if there is, which could lead to an RF loss.

The antenna used in [17] and in GUSTO [18] was modified to accommodate the two-times wider bridge at its center. The log spiral (defined by $R = R_0 \, exp \, (\alpha. \, \phi)$, where $R$ is the distance

This work was supported partly by NWO (project no. 614.061.609) and partly by TU Delft Space Institute (Corresponding author: Behnam Mirzaei; Jian Rong Gao).

B. Mirzaei, D. Ren and J. R. Gao are with the Department of Imaging Physics at Delft University of Technology, Delft, The Netherlands (emails: bmmi.644@gmail.com, D.Ren-1@tudelft.nl, j.r.gao@sron.nl).

J. R. G. Silva, W. J. Vreeling, W. Laauwen, and J. R. Gao are with SRON Netherlands Institute for Space Research, Leiden/Groningen, The Netherlands (emails: j.r.g.d.silva@sron.nl & W.J.Vreeling@sron.nl & W.M.Laauwen@sron.nl).

from the center, $R_0$ is the inner radius, and $\phi$ is the azimuthal angle) was adapted and optimized in COMSOL with the following parameters for best coupling to the lower frequencies: $R_0 = 4$ µm and $\alpha = 0.39$. The device is fabricated on a 380 µm thick, highly resistive Silicon wafer covered by a ~ 5 nm sputtered NbN film [19] with a $T_c$ of 9.9 K.

The spiral antenna including contact structures was patterned by evaporation of a thick, 200 nm Au on the NbN film and a lift-off. The crucial step of properly in-situ Ar$^+$ milling (cleaning) of the film prior to Au evaporation was performed to realize the Au/NbN interface as transparent as possible. The optimal milling parameters were determined by extensive pre-tests on a similar NbN film. In detail, the milling beam voltage is 350 V. The beam current is 38 mA. The acceleration voltage is 600V. The discharge voltage is 30 V. The sample was rotated by 10 rpm. The etching duration is 50 sec. It etches the native oxide (assumed) and also about 8% of the NbN film, corresponding to 0.4 nm, which is monitored by the change of the sheet resistance of the film.

In contrast to the method of adding a superconducting interlayer (e.g., Nb or NbTiN) between the Au contacts and NbN [15,16] e.g., for GUSTO's detectors, we here directly deposit the thick Au (200 nm) on the cleaned NbN, which is expected to provide better contacts and thus a lower loss of the RF current from the antenna to the NbN bridge. Besides, this process also reduces the number of fabrication steps, making the fabrication more reliable and reproducible. A similar approach for the contact structures was reported earlier in [20]. An earlier work of HEBs, whose Au/NbN contacts are deposited in-situ, has shown an extremely low $T_{rec}^{DSB}$ of 600 K at 2.5 THz, suggesting also the importance of the interface [21].

In the next step a sub-µm NbN bridge area is masked by a negative E-beam resist followed by the reactive ion etching of the rest of the surface till the remaining bare NbN is fully removed. An optical micrography of such a detector is shown in the inset of Fig. 1.

We measured the resistance vs temperature (RT) of several HEBs with a length of either 200 or 400 nm. As shown in Fig.1, a $T_c$ of ~ 8.8 K for the 400 nm long HEB was typically found, which is about 1 K lower than the $T_c$ of the original film. Furthermore, from the single transition feature in the RT and non-zero resistance at 4 K, the contact regions of Au/NbN become fully normal, confirmed through a dedicated test structure. We also found a $T_c$ as low as 8.2 K obtained from an HEB with a 200 nm long bridge. This confirms a (lateral) proximity effect, inducing from the thick Au contacts to the superconducting NbN bridge, and weakening its superconductivity. This suggests that we have a transmissive (or clean) interface between Au and NbN in the contacts [12].

### III. EXPERIMENTAL SETUP

We applied a heterodyne measurement setup in air to measure receiver noise temperature, which is very similar to what reported in [14, 22]. In detail, a far-infrared (FIR) gas laser generates the LO signal at three frequencies of 1.63, 2.52, and 5.25 THz, which are quoted as 1.6, 2.5, and 5.3 THz respectively in this paper. These frequencies are very close to what are required for SALTUS's mixers. The amount of the LO power to the HEB is controlled by a swing-arm voice coil actuator, where the swing-arm blocks a part of the LO radiation, and thus acts as an effective variable optical attenuator [23], located in front of the FIR laser. The HEB mixer is mounted at the back of a 10 mm diameter elliptical silicon lens, with an antireflection coating optimized for 1.6 THz [24], which couples the radiation to the antenna. The mixer block is placed in a liquid helium cryostat. The THz signal and LO pass through a UHMW-PE window at 300 K and a heat filter (QMC) at 4 K, with a cut-off frequency of 5.8 THz. A 3 µm thick Mylar beam splitter is used to combine the radiation of a hot or cold (295/77 K) load with that of the LO. The hot and cold loads are needed for the Y-factor measurements to determine $T_{rec}^{DSB}$. The mixer is biased and read out using a bias-T followed by a cryogenic low noise SiGe amplifier (LNA) inside the cryostat with a noise temperature ($T_{IF}$) of 6.5 K. Between the bias-T and LNA there is a circulator to prevent standing waves between the LNA and the HEB. The IF signal is filtered at 1.7 GHz within a band of 80 MHz and is further amplified by a room temperature LNA before it is measured by a power meter. The mixer block is at a temperature of 4.6 K.

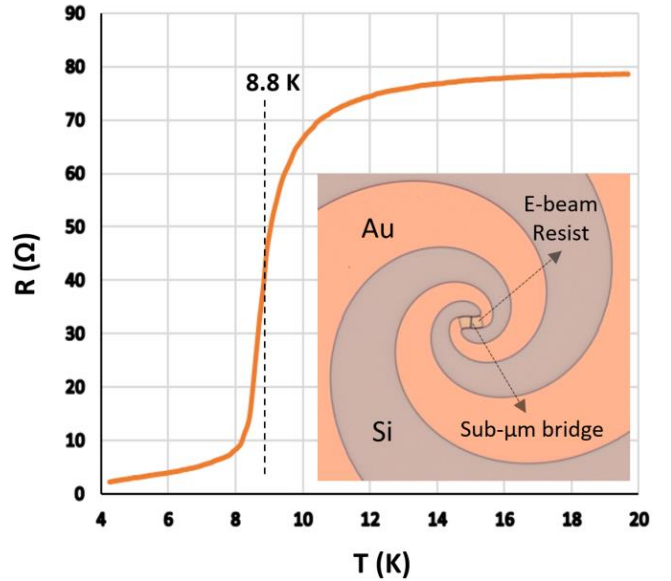

Fig. 1. Resistance versus temperature around the superconducting transition of a 400 nm (length) × 4 µm (width) NbN HEB with cleaned Au/NbN contacts. Inset: optical micrograph of the device showing the spiral arms around the NbN bridge.

The optical losses (L) of the components in the optical path from the hot/cold load to the HEB antenna are summarized in Table I. The air path is ~ 30 cm long with a relative humidity of ~ 40 %. It is worthy to mention that the coupling loss between the antenna and the HEB [17] is neither listed in this table nor included in the calculation of $T_{mixer}^{DSB}$ and the mixer conversion loss $L_{mixer}^{DSB}$ from the measured $T_{rec}^{DSB}$ and receiver conversion loss $L_{rec}^{DSB}$. The reason for this will be discussed in section IV.

TABLE I
OPTICAL LOSSES CREATED BY AIR, 3 µm BEAM SPLITTER,

UHMW-PE WINDOW, QMC HEAT FILTER, AND SILICON LENS WITH ANTIREFLECTION COATING AT 1.6 THZ. THE LOSS OF THE LENS AT EACH FREQUENCY WAS SIMULATED USING COMSOL

| LO Frequency | $L_{air}$ (dB) | $L_{BS}$ (dB) | $L_{window}$ (dB) | $L_{filter}$ (dB) | $L_{lens}$ (dB) |
|---|---|---|---|---|---|
| 1.63 THz | 0.64 | 0.09 | 0.38 | 1.14 | 0.35 |
| 2.52 THz | 0.38 | 0.19 | 0.61 | 0.34 | 1.11 |
| 5.25 THz | 0.90 | 0.63 | 1.47 | 0.56 | 1.14 |

## IV. RESULTS AND DISCUSSION

We focus only on the NbN HEB mixer labeled as "OASIS BM2 7B" although very similar performance was found for two other HEBs from the same batch (wafer) and another. A set of measured current-voltage (IV) curves without and with different LO pumping levels at 1.6 THz are shown in Fig. 2, where the optimum operating point resulting in the best $T_{rec}^{DSB}$ is indicated.

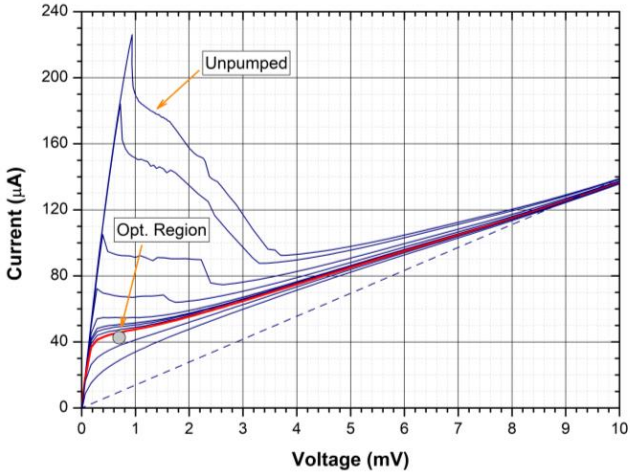

Fig. 2. Unpumped and pumped IV curves of the NbN HEB at 1.6 THz. The thick, red curve is the optimally pumped IV. The optimum operating region within roughly 5 % of the lowest $T_{rec}^{DSB}$ is indicated with a circle, which covers two bias points of 0.6 and 0.8 mV. The dashed load line is used to estimate LO power at HEB itself.

The $T_{rec}^{DSB}$ of the mixer is obtained using the Y-factor that is the ratio between the two measured receiver output powers responding to the hot ($P_{hot}$) and the cold loads ($P_{cold}$). By combining the Y-factor and the Callen-Welton temperatures of the blackbody loads, the $T_{rec}^{DSB}$ can be obtained [25]. The $P_{hot}$ and $P_{cold}$ are recorded as a function of HEB current at an optimal DC bias voltage (0.6 mV), where the current variation reflects the scan of LO power by the voice coil attenuator. In this way, the measurement of a $T_{rec}^{DSB}$ is not influenced by the fluctuations and drift of the FIR laser power [25]. The latter are known for FIR gas lasers. Furthermore, this technique using the same-current to determine the Y-factor can mitigate the direct detection effect [26-29] on it and thus on $T_{rec}^{DSB}$. In essence, the same current is a result of slight change of LO power ($P_{LO}$) to compensate the difference between the broad band hot and cold load power. As a result, the bias points in two cases are adjusted to be the same, but $P_{LO}$ is no longer the same. So, conceptually this is not fully correct because of a difference in the mixer gain [30, 31]. However, in the case of practical NbN HEBs in this work, where the change of $P_{LO}$ is small and is only 2-4 % of the required $P_{LO}$, this method can mitigate the direct detection effect effectively without overestimating $T_{rec}^{DSB}$. This conclusion was based on our earlier experiments in [29, 25], where we made a comparison between the same-current method and the conventional fixed-LO-power-method, where a cold, narrow bandpass filter (CNBPF) centered at the LO frequency was introduced. Although the same current-method was used in [26, 28, 32] and in many of our own publications, ideally, we should apply a CNBPF [25, 27, 28] to mitigate the direct detection effect. Unfortunately, we do not have such CNBPFs for three different frequencies available. It is interesting to note that double-current method measures a slightly lower (~10 %) $T_{rec}^{DSB}$ than the conventional fixed-LO-power-method even for a waveguide HEB (without an CNBPF) because the latter still suffers from the direct detection effect [32]. The receiver output power (IF power) measured in response to the hot/cold loads together with the calculated $T_{rec}^{DSB}$ are shown for all three frequencies in Fig. 3. The minimum $T_{rec}^{DSB}$, taken at a current of around 45 µA, is 530, 640, and 2180 K at 1.6, 2.5, and 5.3 THz, respectively. We have repeated measurements by varying the voltage until 1.2 mV in a step of 0.2 mV. We obtain very similar minimum $T_{rec}^{DSB}$ values at 0.8 mV, but with a slightly higher current.

We determine the receiver conversion loss ($L_{rec}^{DSB}$) or total loss of the mixer at similar operating points using a standard U-factor technique [33]. The $L_{rec}^{DSB}$ is 6.6, 7.7, and 11.7 dB at 1.6, 2.5, and 5.3 THz, respectively.

We also estimate the absorbed LO power at HEB from the optimally pumped IV curve using the isothermal technique [34]. The $P_{LO}$ is around 240 nW for all three frequencies, which seems to be a good compromise between availability of the LO sources and stability of the receiver, where a too high requirement of LO can challenge the LO source, while a too low requirement of LO can cause the HEB to be sensitive to any fluctuation of THz power [23]. It is also useful to stress that we will operate the HEB at a current of ≤ 45 µA for the given voltage (e.g., 0.6 mV). The reason for this is to be discussed. Table II summarizes all the measured receiver performance parameters of $T_{rec}^{DSB}$, $L_{rec}^{DSB}$, and LO power at three frequencies.

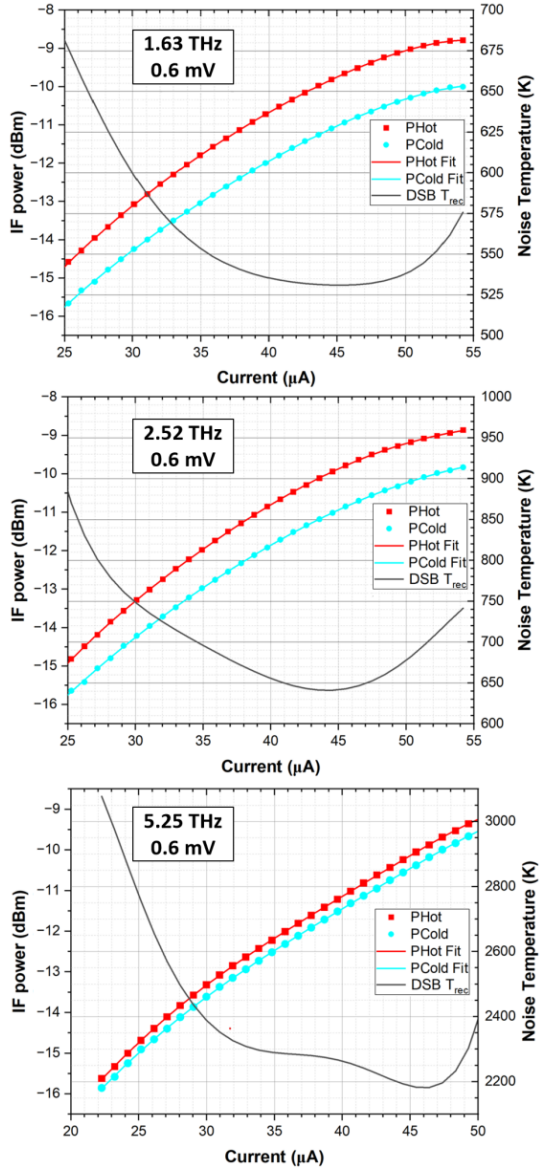

Fig. 3. Receiver output power (IF power) in response to hot/cold loads and the fitted curves (left axis), and receiver noise temperature $T_{rec}^{DSB}$ (right axis) versus current at three LO frequencies 1.6, 2.5, and 5.3 THz and bias voltage of 0.6 mV.

TABLE II
MEASURED DOUBLE SIDEBAND (DSB) RECEIVER NOISE TEMPERATURES ($T_{rec}^{DSB}$), DSB RECEIVER CONVERSION LOSSES ($L_{rec}^{DSB}$), AND ABSORBED LO POWERS

| LO frequency | $T_{rec}^{DSB}$ | $L_{rec}^{DSB}$ | LO power |
|---|---|---|---|
| 1.63 THz | 530 K | 6.6 dB | 245 nW |
| 2.52 THz | 640 K | 7.7 dB | 235 nW |
| 5.25 THz | 2180 K | 11.7 dB | 233 nW |

To be able to fairly judge the mixer performance independent of the measurement setup, we extract $T_{mixer}^{DSB}$ and mixer conversion loss $L_{mixer}^{DSB}$ from the measured receiver parameters in Table II by taking advantage of Table I in the same way as [35]. In detail, for 1.6 THz we subtract the first four optical loss terms listed in Table I, but not the loss term of the coated Si lens, leading to a $L_{mixer}^{DSB}$. Then we apply Eq.1 in [35] using the noise due to the optics ($T_{OP}$) being 101 K, and $T_{IF}$ = 6.5 K to derive the $T_{mixer}^{DSB}$. It is worthy to stress that our $L_{mixer}^{DSB}$ includes both the optical loss of the antireflection coated Si lens (0.35 dB) and the antenna-HEB coupling loss (not known).

TABLE III
DERIVED DOUBLE SIDEBAND (DSB) MIXER NOISE TEMPERATURE $T_{mixer}^{DSB}$ (INCLUDING THE CONTRIBUTION OF THE SI LENS IDEALLY ANTIREFLECTION COATED FOR EACH SPECIFIC FREQUENCY AND ANTENNA-HEB COUPLING LOSS), $T_{mixer}^{DSB}$ IN UNIT OF $h\nu/k_B$ (QN), AND DSB MIXER CONVERSION LOSS ($L_{mixer}^{DSB}$). NOTE THAT ALL ARE OBTAINED AT AN IF OF 1.7 GHZ AND AN OPERATING TEMPERATURE OF 4.6 K

| LO frequency | $T_{mixer}^{DSB}$ | $T_{mixer}^{DSB}$ (Quantum noise unit) | $L_{mixer}^{DSB}$ |
|---|---|---|---|
| 1.63 THz | 240 K | 3.1 × $h\nu/k_B$ | 4.4 dB |
| 2.52 THz | 290 K | 2.4 × $h\nu/k_B$ | 5.5 dB |
| 5.25 THz | 620 K | 2.5 × $h\nu/k_B$ | 7.4 dB |

To drive $T_{mixer}^{DSB}$ and $L_{mixer}^{DSB}$ at 2.5 and 5.3 THz, we do slightly differently using the measured data in Table II and the optical loss in Table I, but assuming that the lens in both cases is optimally coated for each frequency so that the loss is about 0.35 dB, similar to 1.6 THz. So, by using $T_{OP}$ of 116 K and 401 K for 2.5 and 5.3 THz, respectively, we derive both $T_{mixer}^{DSB}$ and $L_{mixer}^{DSB}$. Like in the case of 1.6 THz, here the $L_{mixer}^{DSB}$ include the contribution of the antenna-HEB coupling loss. Table III summarizes the mixer performance parameters i.e., $T_{mixer}^{DSB}$ and $L_{mixer}^{DSB}$, which are likely the lowest values ever reported for HEBs. To get a feeling how close the $T_{mixer}^{DSB}$ is to the fundamental noise limit, we also quote it in units of the quantum noise (QN, $h\nu/k_B$) [9] in this table, where $h$ is the Planck constant, $\nu$ frequency, and $k_B$ the Boltzmann constant. We consider Table III to be the key outcome of this paper.

To better sense the improvement we have made in THz HEBs, we will compare with $T_{mixer}^{DSB}$ and $L_{mixer}^{DSB}$ reported by others or by our team if they are available, otherwise we will compare $T_{rec}^{DSB}$ only. We now compare $T_{mixer}^{DSB}$ and $L_{mixer}^{DSB}$ with what we obtained from the HEBs used for GUSTO arrays [14], where they are spiral antenna coupled NbN HEBs of 0.2 μm × 2 μm in size with the contact structures of Nb/Au bilayer. More than 16 HEBs were tested at 1.6 THz and similar numbers were tested at 5.3 THz (from which we extrapolated the 4.7 THz data for GUSTO).

It is useful to note that at 1.6 THz the GUSTO test setup was the same as what was used in this work i.e., all optical losses are similar. However, all the data were measured at an IF of 2 GHz, so we extrapolate the data to 1.7 GHz for a comparison. For GUSTO HEBs we found the average $T_{mixer}^{DSB}$ and $L_{mixer}^{DSB}$ of 390 K and 6.7 dB, respectively. Comparing with the $T_{mixer}^{DSB}$ in table III (240 K), our new result has an improvement of 39 %.

This can reduce the integration time of a heterodyne observation instrument by a factor of 2 if we conservatively take only about 30 % since the integration time is proportional to the square of the noise temperature [9], which is remarkable. In other words, the observation of a telescope can be nearly 2 times faster, which significantly impacts the determination of the lifetime of a space mission.

Not only the $T_{mixer}^{DSB}$ is improved, but also the $L_{mixer}^{DSB}$ in Table III (1.6 THz) is roughly 2 dB lower than what found for GUSTO HEBs. The latter can reduce the noise contribution of an LNA (referred to Eq.1 in [35]) and thus a receiver (or system) noise temperature of a practical instrument.

No data at 2.5 THz from the HEBs for GUSTO were available to compare. But there are numerous $T_{rec}^{DSB}$ data at this frequency published in the literature. The $T_{rec}^{DSB}$ of 600 to 900 K, measured within 1-2 GHz at IF, was reported in the HEB with in-situ made Au/NbN contacts in [21], where the lowest value is comparable to our 2.5 THz data. Other examples include a $T_{rec}^{DSB}$ of 780 K reported in [36], which used also Ar$^+$ cleaned Au/NbN contacts, 900 K in [37], 1400 K in [38], and 630 K in [23], all obtained using the same setup but in vacuum [25] at SRON, that means less optical loss than our air measurement setup discussed here e.g., no loss of air and the window. Furthermore, 1100 K using an air setup was reported in [39], 950 K in [40], and 1800 K in [41]. The latter was measured earlier in a Nb diffusion-cooled HEB. So, we can generally conclude that our new $T_{rec}^{DSB}$ at 2.5 THz is at least 30 % lower than what reported so far in the literature [36-41]. Here we compare $T_{rec}^{DSB}$ instead of $T_{mixer}^{DSB}$ because the latter is often not available in the literature or is unable to be derived due to some missing information.

Less improvement of the $T_{mixer}^{DSB}$ at 5.3 THz may be partly due to the QN, which plays a more vital role when the frequency increases [42].

At 5.3 THz, the HEBs for GUSTO have shown the average $T_{mixer}^{DSB}$ and $L_{mixer}^{DSB}$ of 760 K and 9.3 dB at an IF of 1.7 GHz [14], respectively. The new $T_{mixer}^{DSB}$ (620 K) shows 18 % improvement, which is marginal. We also compare our new result of $T_{rec}^{DSB}$ (2180 K) at 5.3 THz in the literature, where we find it comparable with what reported in [17] and [43] using the vacuum setup at SRON, by correcting for the difference in the optical loss. In these cases, there is almost no improvement.

There are more measured performance data of HEBs in the literature. However, due to the difference in the LO frequency, we do not make a detailed comparison here, but quoting some recent results of measured $T_{rec}^{DSB}$: 700 K at 1.9 THz (at 1.7 GHz IF) [36] and 950 K at 4.75 THz (at IF between 1.25-1.75 GHz) [32].

Here we try to address possible questions that might arise from the measured data. First, why the $L_{mixer}^{DSB}$ in Table III increases by 3 dB from 1.6 THz to 5.3 THz? We believe this is due to an increase in coupling loss between the antenna and the HEB. This is caused by the antenna modification to accommodate a 4 μm wide NbN bridge, leading to a less tight winding than the antenna used in [17,18] with an inner diameter of 6.6 μm, which well performs at higher frequencies. In contrast, the antenna here has an inner diameter of 12 μm, being less favorable for 5.3 THz [45]. To discuss this quantitatively, we simulate the coupling loss between the antenna and HEB for our case in the same way as in [17], but using COMSOL. We find the coupling loss of 0.13 dB (the efficiency of 97 %), 0.36 dB (92%), and 1.6 dB (69%) for 1.6, 2.5, and 5.3 THz, respectively. Clearly the coupling loss at 5.3 THz is more than that (0.63 dB) for the antenna used in [17,18]. This can explain a large part of the increase in $L_{mixer}^{DSB}$ in table III. To further quantify the discussion, we calculate the intrinsic DSB mixer conversion loss $L_{mixer}^{DSB,intrin}$ using the data of $L_{mixer}^{DSB}$ from table III by subtracting the antenna-HEB coupling losses and also the loss of 0.35 dB for coated Si lenses at three frequencies. We find the $L_{mixer}^{DSB,intrin}$ is 3.9, 4.8, and 5.4 dB for 1.6, 2.5, and 5.3 THz, respectively. These values are considerably lower than what was reported in [42] and lower than others reported in THz HEBs as far as we know. However, the remaining increase in the $L_{mixer}^{DSB,intrin}$, observed also in [42], is not expected.

We expect the $L_{mixer}^{DSB,intrin}$ to be frequency independent in the frequency range of our experiment, where THz photon energies are considerably higher than the gap frequency of the NbN bridge. This is supported by the close-to-optimally pumped IV curves at three frequencies being extremely well overlapped, shown in Fig. 4. This result suggests that the effect of the heating of THz current in the bridge is the same at the three frequencies. We can further speculate that despite of the wide NbN HEB (4 μm), the THz current flow along the bridge is likely uniform, as found for a narrower NbN HEB (2 μm) [39]. This is in particular interesting in view of the fact that the skin depth for the NbN at 5.3 THz is 0.43 μm, which is only ~ 10 % of the bridge width. As a byproduct of this discussion, we can state that it is still safe to apply the normal state resistance as the RF impedance of the 4 μm HEB at a high frequency of 5.3 THz.

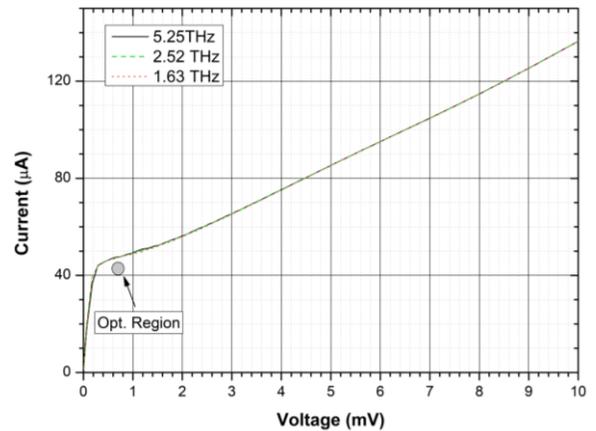

Fig. 4. Close-to-optimally pumped IV of the NbN HEB at three frequencies of 1.6, 2.5, and 5.3 THz. They seem to be fully overlapped.

The strong lateral proximity effect observed on the RT curve suggests the transparent or clean contacts in our HEBs. Such contacts can reduce the RF loss, as we discussed in the introduction. Besides, they can also affect the temperature profile of the hot electrons across the bridge as well as in the contact regions. The measured low mixer conversion losses at 1.6 and 2.5 THz in table III seem to support these hypotheses. Following a hot spot model of mixing for an HEB [31, 47], the contact properties are expected to affect the conversion loss and noise qualitatively. Our improved $T_{mixer}^{DSB}$ confirms that the contacts play a crucial role experimentally, but not yet theoretically. The reported simulations, such as in [31,48] based on the hot-spot model are unable to address this effect since most of them do not include the role of the contacts.

Our contacts behave differently from the earlier Au/NbN contacts [15, 40], which do not change the $T_c$ of the contact region and Nb bridge, suggesting poor interfaces as a result of no $Ar^+$ cleaning.

The enhanced sensitivity of the NbN HEB mixer by engineering the contacts reported here can also stimulate the development of novel $MgB_2$ HEB mixers [35,43,44], which can offer a high operating temperature of ≥20 K and a large IF bandwidth of ≥10 GHz. However, little has been explored on the contacts [21].

The optimal operating region of the HEB is indicated in Fig. 2. More specifically, at 0.6 mV we will choose the HEB to be operated at the current of 45 µA or below as suggested by Fig. 3. One should avoid working at a current above 45 µA because it moves away from the resistive state, where a HEB is known to be unstable [51] and where relaxation oscillations switching between the superconducting state and resistive state are present [52]. Such oscillations were also recently reported in [53]. The same work also found weaker, intrinsic oscillations when a NbN HEB is in the resistive state without LO power and is operated near its $T_c$. Although it is unclear whether the intrinsic oscillations can affect our performance of the HEB, it might be worthy to perform a dedicated measurement to localize the region of low $T_{rec}^{DSB}$, but being free of any unwanted oscillations.

## V. CONCLUSION

By introducing cleaned thick Au contacts to a NbN HEB mixer we have demonstrated extremely low mixer noise temperatures ($T_{mixer}^{DSB}$) of 240 and 290 K at 1.6 and 2.5 THz, respectively, which are about 3 times the quantum noise (hv/kB). The mixer is developed for future FIR space observatories, in particular, for SALTUS mission concept, the proposal of which will be submitted to NASA in response to the Probe mission call. The improvement of $T_{mixer}^{DSB}$ is more than 30 % in comparison with reported NbN HEB mixers [14]. This can reduce the integration time of a heterodyne instrument roughly by a factor of 2. In other words, such an improvement can make the observation twice as fast, that is highly demanded for cryogenic space missions in view of limited resources. The $T_{mixer}^{DSB}$ of the same HEB at 5.3 THz has shown 620 K and thus a limited improvement, which is partly due to the non-optimized antenna geometry, supported by the high antenna- HEB coupling loss, and is likely also due to the QN at this high frequency. Interestingly, despite of a 4 µm wide NbN bridge, the pumped IV curves are identical at three different frequencies, suggesting that the RF current flows uniformly along the bridge.

## ACKNOWLEDGMENT

The authors thank Yuner Gan for modeling the loss of the coated Si lens. We acknowledge Marc Zuiddam and the staff in Kavli Nanolab Delft for their technical support. We are also grateful for the huge support from Paul Urbach at TU Delft, and Pieter Dieleman, Jan-Willem den Herder, and Jan Geralt bij de Vaate at SRON.